# Fast Neutron Detection Efficiency of $^6$Li and $^7$Li Enriched CLYC Scintillators using an Am-Be source

N. Blasi[1], S. Brambilla[2], F. Camera[1,2,*], S. Ceruti[1,2], A. Giaz[1,2,†], L. Gini[2], F. Groppi[1,2], S.Manenti[2], A. Mentana[1,2], B. Million[2], S. Riboldi[1,2]

[1] *University of Milano, Department of Physics, Via Celoria 16, 20133 Milano, Italy*
[2] *Istituto Nazionale di Fisica Nucleare (INFN), Sezione di Milano, Via Celoria 16, 20133 Milano, Italy*

ABSTRACT: The fast neutrons produced by a calibrated $^{241}$Am-Be source were detected by two different $Cs_2LiYCl_6$:Ce (CLYC) scintillator detectors. The two cylindrical crystals (1"x1" in size) were enriched with more than 99% of $^7$Li (C7LYC) and with about 95% of $^6$Li (C6LYC), respectively. Both crystals can detect fast neutrons whereas only C6LYC can also detect thermal neutrons, due to the presence of $^6$Li. The measurement was performed at the L.A.S.A. Laboratory of INFN and University of Milano (Italy). To identify the neutron events, the Pulse-Shape-Discrimination technique was used. A value for the detection efficiency of the $^{241}$Am-Be emitted neutrons, with energy up to 10 MeV, was deduced.

[*] *corresponding author:* franco.camera@mi.infn.it
[†] *present address: University of Padova, Department of Physics, via Marzolo 8, Padova, Italy*

________________________________________________________________

Contents

________________________________________________________________

## 1. Introduction

The interest in elpasolite crystals, and in particular in Cerium doped $Cs_2LiYCl_6$ (CLYC) [1] arises from the fact that they can detect not only gamma rays with an energy resolution of ~4% at 662 keV, but also thermal and fast neutrons [2-8]. In particular, thermal neutrons are detected with high efficiency via the reaction $^6Li(n,t)\alpha$, which has a cross section of 940 b, and produce a high light output at Gamma Equivalent Energy (GEE) of ~3.3 MeV. In this paper, we will focus on fast neutrons, and, in particular, on neutrons in the energy range 0.8-10 MeV. These neutrons are detected mainly via the reactions $^{35}Cl(n,p)^{35}S$ and $^{35}Cl(n,\alpha)^{32}P$ with cross sections of the order of $10^{-1}$ b. In this case, the energy of the proton or the $\alpha$ particle is directly related to the neutron energy [8-11]. This simple relation can be taken advantage of up to neutron energies of about 5 MeV [5, 10, 11, 12]. Above that energy, too many reaction channels are open, resulting in a spreading of the response function over the whole spectrum. Separation between γ-rays and neutrons can be made by pulse shape discrimination (PSD). Several PSD ratios or figures of merit (FOM) are reported in literature [5, 6, 7, 13]. For example, a FOM typical value of 2.7 corresponds to a γ-ray rejection ratio better than $10^{-7}$ [13]. In general, thermal neutrons dominate the CLYC neutron response function and only Time of Flight (ToF) measurements can disentangle them from fast neutrons. However, since thermal neutrons react with $^6Li$ but not with $^7Li$, by using $^7Li$ enriched ingots one obtains crystals (C7LYC) which are only sensitive to fast neutrons, as for example shown in reference [5].

This unique feature of detecting simultaneously γ-rays and fast neutrons makes this crystal extremely interesting for nuclear physics experiments. For that purpose, however, it would be important to have a direct measurement of the neutron detection efficiency. The thermal neutron efficiency for CLYC was discussed, for example, in reference [14]. Due to the difficulty to perform such a measurement on fast neutrons, not many studies are available in literature. Allwork et al. [15] used a $^{252}Cf$ source on a C6LYC of dimension 2"x2" and reported a measured efficiency of 0.99 ±0.01 % for neutrons ranging from 2 to 6 MeV. Soundara-Pandian et al. [16] measured the intrinsic fast neutron efficiency of a $^{252}$Cf source on a CLYC as a function of the crystal dimensions and concluded that the efficiency doubles going from 1"x1" to 3"x3". Smith et al. [9] used a 1"x1" CLYC and four mono energetic neutron beams from 4.1 to 5.5 MeV. They considered only the neutron events populating the ground state of $^{35}$S via the $^{35}$Cl(n,p)$^{35}$S reaction and obtained an efficiency ranging from 0.3% at beam energy 4.1 MeV to 0.16% at 5.5 MeV. D'Olympia et al. [10] performed several measurements with mono energetic neutrons ranging from 0.5 to 20 MeV on a 1"x1" C7LYC. Efficiencies were calculated via a full MCNPX simulation, and the obtained values increased with neutron energy from 0.1% at 0.5 MeV to 1.47% at 5 MeV, 2.86% at 10 MeV and 3.29% at 20 MeV. These values seem in reasonable agreement with what was found by Allwork et al. [15]. However, a comparison with the work of Smith et al. [9] on the $^{35}$Cl(n,p)$^{35}$S(g.s.) reaction channel only shows a disagreement by a factor two. In this paper we report a new fast neutron efficiency measurement performed using two 1"x1" CLYC detectors and a calibrated Am-Be source. One of the two detectors was enriched with $^{6}$Li (C6LYC), while the other was enriched with $^{7}$Li (C7LYC). In section 2, we present the experimental setup, the PSD technique and the neutron experimental spectra. In section 3, the reaction mechanisms involved in the neutron detection in C6LYC and C7LYC are discussed, and the expected neutron spectra are compared to the experimental ones. The measured fast neutron efficiencies are reported in section 4. A discussion on the results is reported in section 5 and conclusions are drawn in section 6.

## 2. The measurement

Two 1"x1" cylindrical samples of CLYC scintillators from the RMD Company [17] were used, one enriched with more than 99% of $^{7}$Li (C7LYC) and the other enriched with about 95% of $^{6}$Li (C6LYC). According to RMD, the actual volumes of the crystals are 12.4 cc for C6LYC and 12.1 cc for C7LYC. However, we observed a reduced efficiency for C7LYC. An accurate inspection of the crystal performed by RMD has revealed a partial damage probably due a microscopical leakage in the sealing. The actual active volume resulted to be 10.4 cc. We therefore normalized our results to the active volumes of the crystals.

The measurement was performed at L.A.S.A. Laboratory of I.N.F.N. and University of Milano. Neutrons were produced by a calibrated $^{241}$Am-Be source of nominal activity 1 Ci, corresponding to an emission of $2.4 \times 10^{6}$ neutrons per second. This source generates neutrons of energy up to ~10 MeV [18] through the following nuclear reactions:

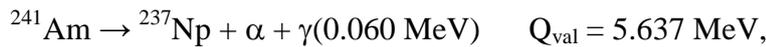

$^{241}$Am → $^{237}$Np + α + γ(0.060 MeV)    $Q_{val}$ = 5.637 MeV,

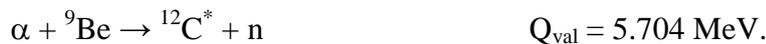

α + $^{9}$Be → $^{12}$C$^{*}$ + n           $Q_{val}$ = 5.704 MeV.

Figure 1 shows the neutron spectrum of a $^{241}$Am-Be source [18].

Fast neutrons are mainly detected via the reactions $^{35}$Cl(n,p)$^{35}$S and $^{35}$Cl(n,α)$^{32}$P with cross sections of the order of $10^{-1}$ b for neutrons of few MeV's. Thermal neutrons are detected with high efficiency via the reaction $^{6}$Li(n,t)α, which has a cross section of 940 b, and produce a high light output at Gamma-Equivalent Energy (GEE) of ~3.3 MeV. At neutron energies of our interest, the cross section for this reaction drops to ~ $10^{-1}$ b, comparable with the cross section on $^{35}$Cl.

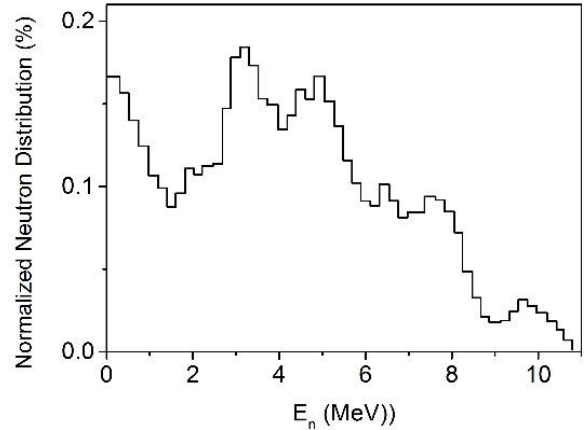

**Figure 1.** Normalized neutron energy distribution (NND) from a $^{241}$Am-Be source. Data are taken from reference [18]

**2.1 The experimental set-up**

In order to perform the Pulse-Shape-Discrimination, a digital acquisition of the signal was used. The two CLYC scintillators were coupled to HAMAMATSU R6233-100Sel photomultiplier tubes (PMT) and to two standard HAMAMATSU voltage dividers (VD), E1198-26 and E1198-27. The PMTs were powered at 800V (CAEN-N1470).

In order to estimate the dead time, a pulser (10 Hz) was employed in the acquisition.

The anode signal was split using a custom module which amplified the two signal by a factor ~ 4. One splitted signal was sent to a 12 bit, 600 MHz LeCroy waverunner HRO 66Zi oscilloscope to be digitized. The acquisition trigger was made by the OR signal between the other splitted signal and the pulser.

The thresholds used during the experiment correspond to γ-ray energy values of ~0.85 MeV and ~1.45 MeV for C6LYC and C7LYC respectively.

Measurements were performed at different distances from the source for both crystals. In particular, C6LYC was placed at d= 0.775, 1.0, 1.2 and 1.4 m and C7LYC at d=1.0 and 1.4 m. Furthermore, a background measurement without source was performed for each crystal.

**2.2 Neutron discrimination with PSD technique**

The pulse digitalization allows to identify neutrons and to separate them from γ-rays by the Pulse-Shape-Discrimination (PSD) technique. The discrimination is performed by comparing the charge integrated over two regions of the pulse [7].

The PSD ratio R is given by the formula:

$$R = \frac{A[W_2]}{A[W_1] + A[W_2]}$$

where $A[W_i]$ is the integral of the signal samples, over the two different integration windows: $W_1$ from the onset to 80 ns, corresponding to the signal rise, and $W_2$ from 100 ns to 600 ns, corresponding to the signal decay. Figure 2 shows the PSD matrix of C6LYC, where we plotted in the x axis the total deposited energy (obtained by integrating the signals from the onset up to 3μs) and in the y axis the PSD ratio, R. Two main event groups can be observed in the matrix: one with $0.90 \leq R < 0.96$, corresponding to neutron events, and a lower one, with $0.80 \leq R < 0.90$, corresponding to γ-ray events.

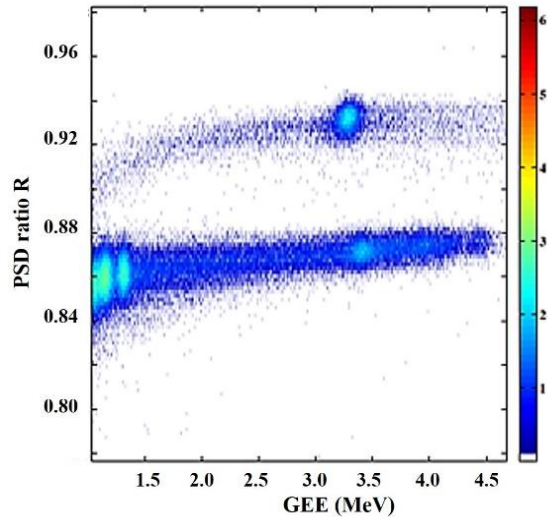

**Figure 2**. A typical PSD matrix of $^{241}$Am-Be and $^{60}$Co sources acquired with C6LYC is shown. The PSD ratio R is plotted against the total deposited energy. Two groups of events can be identified in the matrix: the upper one ($0.90 \leq R < 0.96$) corresponds to neutron events and the lower one ($0.80 \leq R < 0.90$) to γ-ray events.

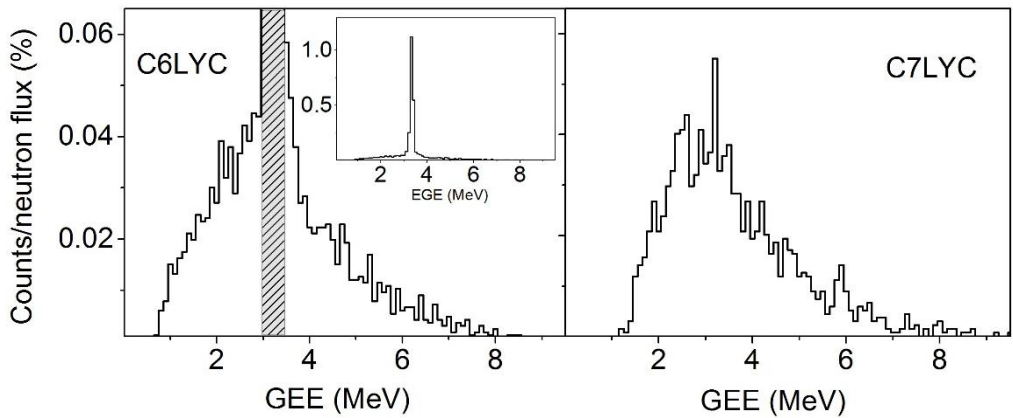

**Figure 3.** Left panel: C6LYC neutron energy spectrum measured at 1.0 m distance. The gray rectangle in the GEE region between 3.0 and 3.5 MeV indicates the thermal energy region. As can be seen in the full scale spectrum shown in the inset, in that energy region the spectrum is dominated by the thermal neutron peak coming from the $^6$Li(n,t)α reaction. Right panel: C7LYC neutron energy spectrum measured at 1.4 m distance. All spectra are normalized to the total neutron flux.

Figure 3 shows the measured neutron energy spectra of C6LYC (left panel) and of C7LYC (right panel). All spectra are normalized to the neutron flux.

As one can see in the left panel, in C6LYC the GEE region around 3.3 MeV is dominated by the thermal neutron peak (see the full scale spectrum in the inset). This area, highlighted in gray in the figure, has been determined as ± 5σ (being σ the standard deviation) around the centroid of the thermal neutron peak of the C6LYC background spectrum. In C7LYC, the contribution of $^6$Li is negligible.

## 3. Expected neutron induced spectrum

As previously stated, the fast neutron detection capability of the CLYC scintillators arises mainly from the presence of $^{35}$Cl, through the reactions $^{35}$Cl(n,p)$^{35}$S and $^{35}$Cl(n,α)$^{32}$P, in which $^{35}$S and $^{32}$P may be left in the ground state or in an excited state. For the sake of simplicity, we will only consider the reaction channels that populate the ground state and the first two excited states of $^{35}$S and $^{32}$P. In C6LYC, the cross section of the $^6$Li(n,t)α reaction rapidly falls as the neutron energy increases to the order of few eV's, but will still give a contribution for neutrons of few MeV's. In this work, the attention is focused on fast neutrons and we will not treat thermal neutrons. The cross sections of the $^{35}$Cl(n,p)$^{35}$S and $^{35}$Cl(n,α)$^{32}$P channels, together with the one

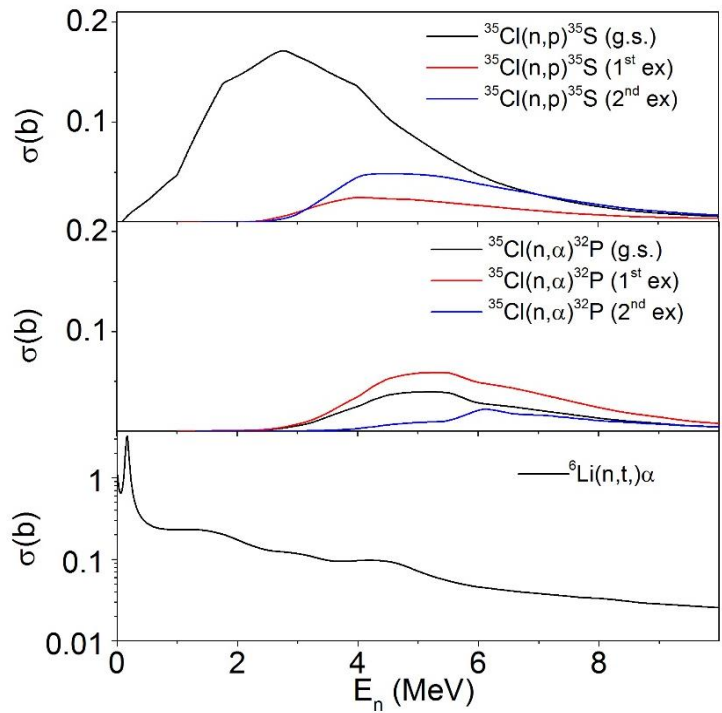

**Figure 4.** Cross sections of the different reaction channels considered [19]. Top panel: $^{35}$Cl(n,p)$^{35}$S with the population of $^{35}$S in the ground state and the first two excited states. Middle

panel: $^{35}$Cl(n,α)$^{32}$P with the population of $^{32}$P in the ground state and the first two excited states. Bottom panel: $^6$Li(n,t)α channel. It is to be noted that the cross section value for this neutron absorption reaction at $E_n \sim 0.025$ eV is about 940 b (not shown).

corresponding to 6Li(n,t)α, are taken from [19] and are shown in figure 4, as a function of the neutron energy En. It should be noted that these calculated cross sections may suffer from large uncertainties, as already pointed out in ref [10].

A neutron with energy $E_n$ may be detected via different reaction channels, each one resulting in a different measured Gamma-Equivalent Energy GEE. In the scintillator, the light output due to the interaction of energetic particles (p, α or t) is proportional to the incident neutron energy $E_n$, corrected by the Q-value of the corresponding reaction channel. Following the notation adopted in [19], we define $Q_M$ as the mass difference Q-value, and $Q_I$ as the Q-value for the excited states of the residual nuclei, $Q_I = Q_M - E^*$. The measured Gamma-Equivalent Energy GEE is expected to be:

$$GEE = (E_n + Q_{M,I}) \cdot f_q. \tag{1}$$

The detector quenching factor $f_q$ depends on the particle type: it is estimated to be ~0.9 for protons, ~0.5 for α particles [6] and ~0.7 for the reaction $^6$Li(n,t)α, where tritons and α particles are both detected. For the sake of clarity, the expected GEE values for three different incident neutron energies $E_n$ are listed in Table 1 for the considered reaction channels. One can see that, for example, a neutron with $E_n = 1$ MeV may be detected at six different GEE values, depending on the particular reaction channel it has undergone, from 0.040 MeV to 4.048 MeV.

**Table 1.** Relation between the neutron incoming energy, $E_n$, and the Gamma-Equivalent Energy (GEE) in MeV, for the seven different reaction channels considered. The $Q_{M,I}$ and $f_q$ values are listed in columns 2 and 3 respectively. The measured values GEE, correspond to a thermal neutron $E_n = 10^{-3}$ MeV, a neutron with energy $E_n = 1$ MeV and one with $E_n = 5$ MeV are listed in columns 4, 5 and 6, respectively.

| Reaction channel | | $Q_{M,I}$ [MeV] | $f_q$ | GEE [MeV] | | |
|---|---|---|---|---|---|---|
| | | | | $E_n = 10^{-3}$ MeV | $E_n = 1$ MeV | $E_n = 5$ MeV |
| $^{35}$Cl(n,p)$^{35}$S | g.s. | 0.615 | 0.9 | 0.554 | 1.453 | 5.035 |
| | 1st exc state | -0.956 | 0.9 | - | 0.040 | 3.640 |
| | 2nd exc state | -1.376 | 0.9 | - | - | 3.262 |
| $^{35}$Cl(n,α)$^{32}$P | g.s. | 0.938 | 0.5 | 0.469 | 0.969 | 2.969 |
| | 1st exc state | 0.860 | 0.5 | 0.430 | 0.930 | 2.930 |
| | 2nd exc state | 0.425 | 0.5 | 0.213 | 0.713 | 2.713 |
| $^6$Li(n,t)α | | 4.783 | 0.7 | 3.349 | 4.048 | 6.848 |

The total energy spectrum is given by the sum of the contributions coming from the different reaction channels. Each contribution is obtained by multiplying the Normalized Neutron energy Distribution probability NND(E) of figure 1 to each reaction channel cross section $\sigma_c(E)$:

$$f_c(E) = NND(E) \cdot \sigma_c(E)$$

We can write the resulting contributions as a function of GEE, using equation (1). In figure 5, the resulting contributions are shown. The total calculated efficiency $\varepsilon_{calc}$ is given by the sum of the contributions of the reaction channels integrated over the whole energy range:

$$\epsilon_{calc} = \sum_c \int_{GEE} K_c \, f_c(GEE) \, dGEE \,,$$

where $K_c = n_c \cdot x$, being $n_c$ the number of atoms of $^{35}$Cl or $^6$Li per volume unit and $x$ the crystal length. For the $^6$Li channel, the thermal peak was spread into a Gaussian with ~ 150 keV FWHM to reproduce the experimental energy resolution. Figure 6 shows the total Am-Be neutron spectra expected for C7LYC and C6LYC. The expected efficiency values for the fast neutrons emitted by an Am-Be source are shown in Table 2 for C6LYC and C7LYC. For C6LYC, the $^6$Li contribution due to thermal neutrons has not been included. The errors take into account the uncertainty of the experimental neutron intensities of figure 1, estimated to be of the order of 4% [18], and systematic errors in $K_c$.

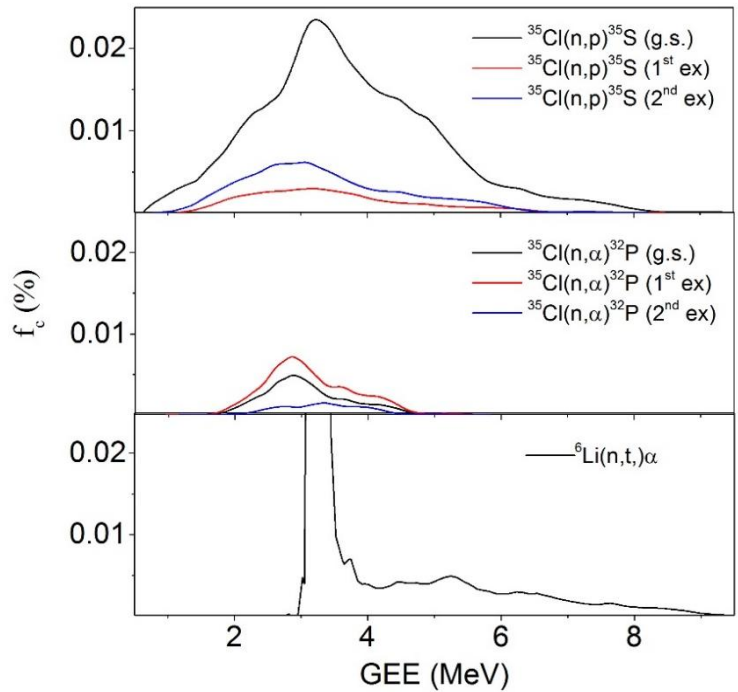

**Figure 5**. Calculated contributions to the total neutron spectrum associated to the different reaction channels, in terms of GEE. Top panel: contributions corresponding to $^{35}Cl(n,p)^{35}S$, central panel: contributions corresponding to $^{35}Cl(n,\alpha)^{32}P$. Bottom panel: contributions corresponding to $^{6}Li(n,t)\alpha$, in which the thermal peak is spread into a gaussian with FWHM of ~150 keV to reproduce the experimental energy resolution.

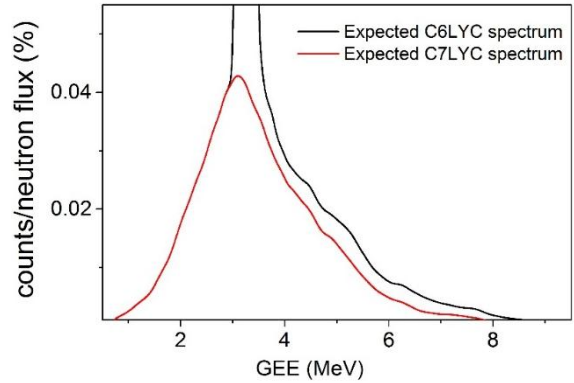

**Figure 6**. Expected spectra of C6LYC (black line), obtained by summing all the seven contributions due to neutron absorption on $^{35}Cl$ and $^{6}Li$, and C7LYC (red line), obtained by summing the six contributions due to neutron absorption on $^{35}Cl$ only.

**Table 2**. Expected fast neutron detection efficiency for an Am-Be neutron emission spectrum.

|       | $\epsilon_{calc}$(%) |
|-------|----------------------|
| C6LYC | 1.23 ± 0.08          |
| C7LYC | 1.07 ± 0.07          |

Figure 7 shows the comparison between the experimental and calculated neutron spectra for C6LYC and C7LYC. One can see that the agreement at energies GEE higher than 3 MeV is rather good, while at lower energies the calculations underestimate the experimental spectra. This is more evident in C6LYC, where the experimental threshold was lower. Possible explanations lie in the several approximations made. In fact, we neglected contributions coming from the population of higher excited states in $^{35}S$ and $^{32}P$, in which the energy of the ejected particle is low. Furthermore, the calculated cross sections we used could be not accurate enough, as already pointed out by D'Olympia et al. [10], and the quenching factors are approximate and might depend on particle energy.

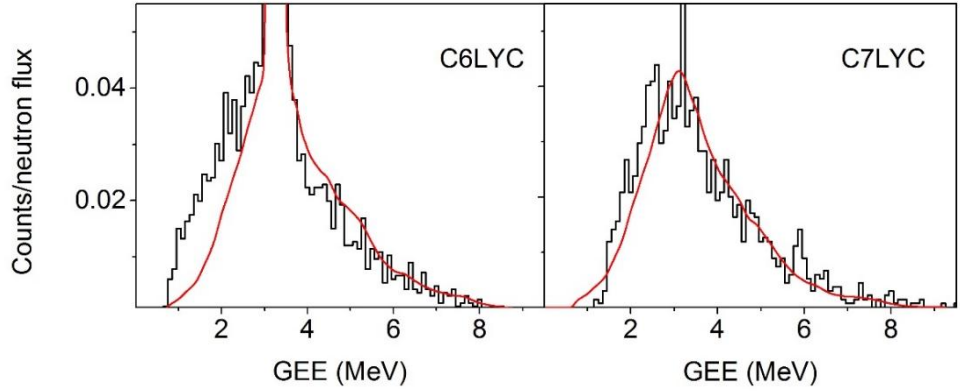

**Figure 7**. Comparison between the measured (black line) and expected (red line) total neutron spectra.

## 4. Fast neutron detection efficiency

The neutron detection efficiency $\epsilon$ is defined as:

$$\epsilon = \frac{\varphi_{meas}}{\varphi_{emit}},$$

where $\varphi_{emit}$ is the neutron flux emitted by the source hitting the detector and $\varphi_{meas}$ is the measured neutron flux.

We will consider four different effective energy regions of the measured energy spectrum. The GEE energy range 0.85-1.45 MeV corresponds to the different experimental threshold energies applied to the measurements, as specified in section 2 (threshold energy region, labelled *thres*). The range 1.45-3.00 MeV corresponds to energies above the threshold region but lower than the thermal peak in C6LYC (low energy region, labelled *low*). The region 3.00-3.45 MeV corresponds to the thermal peak region (thermal energy region, labelled *therm*). The range 3.45-10.0 MeV, covers energies higher than the thermal peak (high energy region, labelled *high*).

The total efficiency for fast neutrons emitted by an Am-Be source will therefore be:

$$\epsilon = S_{thres} + S_{low} + S_{therm} + S_{high}, \qquad (2)$$

where $S_i$ is the integral of the measured energy spectrum over the i-energy region, background subtracted, normalized to the total flux $\varphi_{emit}$ hitting the detector. In Figure 8, the $S_{low}$ and $S_{high}$ values for the two detectors are plotted as a function of the detector distance from the source. As indicated in section 2, in the threshold energy region we have only measurements performed with C6LYC, and in the thermal energy region $S_{therm}$ cannot be determined for C6LYC because of the overwhelming thermal neutron peak. The weighted averages of the $S_k$ values are reported in Table 3.

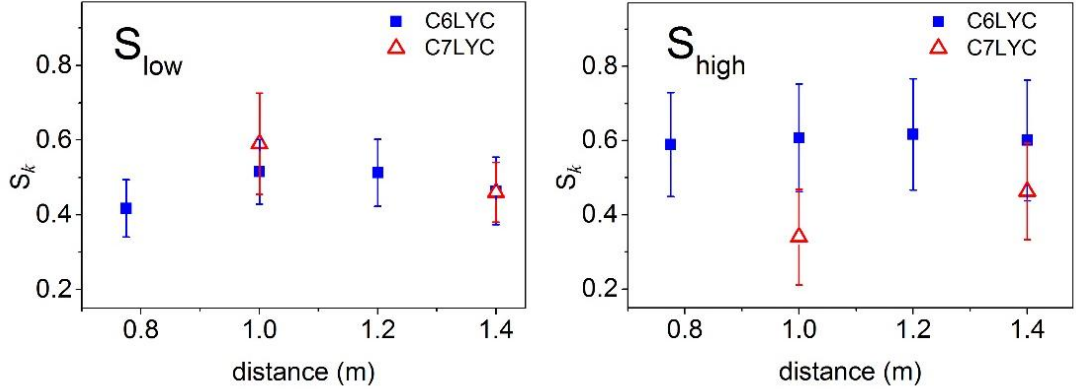

**Figure 8.** $S_{low}$ (left panel) and $S_{high}$ (right panel) for the different source-detector distances for C6LYC and C7LYC.

**Table 3**. $S_k$ values for C6LYC and C7LYC obtained as weighted averages of the measurements at different distances from the source.

|  | $S_{thres}$(%) | $S_{low}$(%) | $S_{therm}$(%) | $S_{high}$(%) |
|---|---|---|---|---|
| C6LYC | 0.07 ± 0.02 | 0.55 ± 0.05 |  | 0.61 ± 0.07 |
| C7LYC |  | 0.50 ± 0.07 | 0.18 ± 0.02 | 0.40 ± 0.09 |

**Table 4.** Measured fast neutron efficiency values of an Am-Be source for C6LYC and C7LYC. The two values refer to the whole energy range considered in this work ($E_{eff}$ =0.85-10 MeVee).

|  | $\epsilon$ (%) |
|---|---|
| C6LYC | 1.41 ± 0.16 |
| C7LYC | 1.16 ± 0.21 |

In the low energy region, the $S_{low}$ values are consistent for C6LYC and C7LYC. In the high energy range, the difference between the two $S_{high}$ values is due to the presence of $^6$Li in C6LYC, that enhances the neutron detection capability (see figure 5).

The fast neutron detection efficiency $\epsilon$ of an Am-Be source can now be determined for each detector by summing all the $S_k$ contributes. In $S_{thres}$ and $S_{low}$ regions, there is no reason to consider C6LYC and C7LYC different, within the error bars, in the detection of fast neutrons. Therefore, the $S_{thres}$ value obtained for C6LYC can be used also for C7LYC. Similarly, in first approximation the $S_{therm}$ value deduced for C7LYC can be used also for C6LYC.

The measured values of the detection efficiency $\epsilon$ for fast neutrons emitted by an Am-Be source are listed in Table 4 for C6LYC and C7LYC. Compared to the calculated efficiencies, listed in table2, the experimental values are ~10% larger, although consistent within the error bars. This discrepancy was however expected since we only took into account few reaction channels in the calculations, and was already observed by comparing the neutron spectra (figure 7).

## 5. Discussion

The efficiency values listed in Tables 2 and 4 refer to neutrons with the energy distribution as shown in figure 1. Since there is no univocal correspondence between the energy of the incident neutron and the energy of the particle detected by the crystal, it is not possible to deduce in a simple way a weighted average of the efficiency for neutrons up to 10 MeV. We can however compare our results to the ones presented by D'Olympia et al. [10], who measured neutron spectra with a C7LYC using monochromatic beams from 0.52 up to 10 MeV, and performed complete Monte Carlo simulations including all possible reaction channels. As can be seen in the left panel of figure 9, the comparison of their simulations with ours shows that the approximation of taking into account the ground state, the first and the second excited states in $^{35}$S and $^{32}$P is valid for neutrons up to ~ 4 MeV.

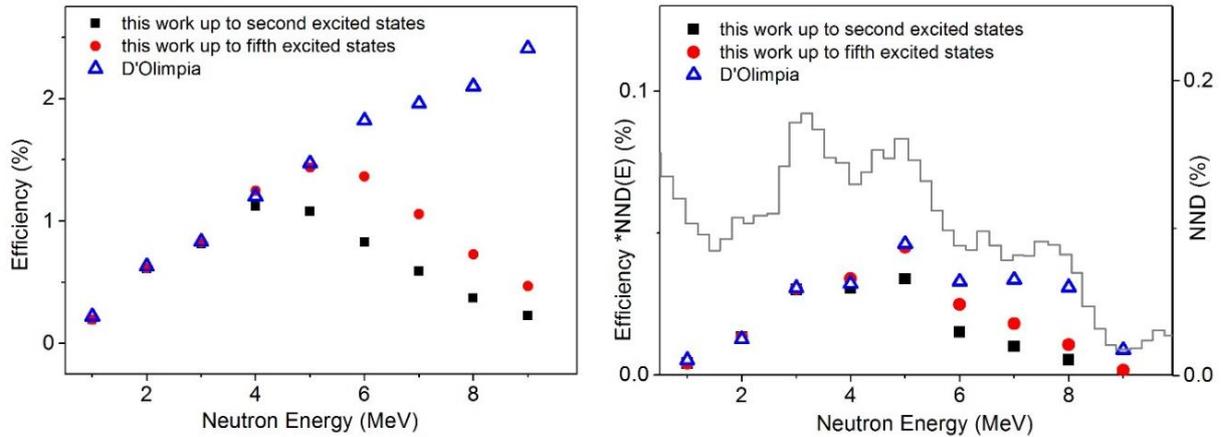

**Figure 9.** Left panel: comparison between the simulation of this work and the complete simulations of D'Olympia et al. [10]. Black squares indicate the efficiency calculated considering the ground state, the first and the second excited states in $^{35}$S and $^{32}$P, as a function of incident neutron energy. Red dots indicate the same simulation including the ground states and the first five excited states. Blue open triangles indicate the simulations of ref [10]. Right panel: the efficiencies of the left panel are multiplied by the Normalized Neutron Distribution (NND(E)). The trend of NND is also plotted (gray line, ordinate scale on the right).

The inclusion of the next three excited states improves the approximation up to ~ 5 MeV. The reaction cross section of higher excited states is not very large, but the number of open reaction channels increases as the neutron energy increases, therefore their contribution is not negligible. However, the contribution of these states most probably falls in the low energy part of the measured spectra, since the charged particle is ejected with low energy. Furthermore, if we consider the energy distribution of the neutrons emitted by the Am-Be source (figure 1 and right panel of figure 9), we see that only ~ 20% of the emitted neutrons has an energy larger than 6 MeV, and the percentage reduces to ~ 7% for neutron energies larger than 8 MeV. When the simulated efficiencies are multiplied by the neutron energy distribution, as shown in the right panel of figure 9, we see that at high energies the discrepancy with the simulations of D'Olympia et al. reduces.

The total efficiency obtained applying the full simulations of D'Olympia et al. to the neutron energy distribution is 1.15% for a C7LYC. By comparing this value with our expected value of 1.07% ± 0.07 (see Table 2) we may conclude that we neglect less than 10% of the total efficiency. Therefore, in first approximation, we can consider our simplified simulation valid. Our measured value of 1.16% ± 0.21 turns out to be in very good agreement with the simulations of D'Olympia et al.

## 6. Conclusions

In this work, the detection efficiency of the fast neutrons produced by an Am-Be source has been measured and estimated for two samples of CLYC scintillator detectors, enriched with $^6$Li (C6LYC) and $^7$Li (C7LYC), respectively. C6LYC is also highly sensitive to thermal neutrons because of the large cross section of the reaction $^6$Li(n,t)$\alpha$. The thermal neutron efficiency for CLYC was discussed, for example, in reference [14] and we did not treat them in this work since the thermal neutron flux was not known.

The experimental neutron spectra were compared to simple simulations where the final nuclei $^{35}$S and $^{32}$P were left in the ground state or in the first or second excited state. It turns out that the calculations well reproduce the spectra for effective measured energies from ~3.3 up to 10 MeV, but underestimate the low energy part of the spectra, because of the small number of reaction channels considered in the calculations. The total measured efficiency to the Am-Be neutrons turns out to be 1.41 ± 0.16 and 1.16 ± 0.21 for C6LYC and C7LYC, respectively. The difference is due to the presence of $^6$Li in C6LYC. The cross section of the $^6$Li(n,t)$\alpha$ reaction is very large for thermal neutrons (~ 940 b), then falls rapidly as the neutron energy increases to the order of few eV's, but will still give a contribution of the order of $10^{-1}$ b for neutrons of few MeV's, comparable to the cross section on $^{35}$Cl. We included the $^6$Li(n,t)$\alpha$ contributions for effective energies larger than 3.45 MeV.

By applying the simulations performed by D'Olympia et al. [10], where all reaction channels were taken into account, to the neutron emission spectrum of an Am-Be source, we find the efficiency value of 1.15 for a C7LYC, which is in good agreement with the value measured in this work. Our results therefore confirm the works of D'Olympia et al. [10] and Allwork et al. [15], and disagree with what was reported by Smith et al. [9].

The forthcoming step will be to perform an in-beam measurement to determine the fast neutron detection efficiency for CLYC crystals of different dimensions (1"x1", 2"x2" and 3"x3") as a function of the neutron energy. Furthermore, we plan to combine the PSD and the time of flight techniques in an attempt to study the relation between the incident neutron energy and the measured effective energy.

## Acknowledgments

We acknowledge Radiation Monitoring Devices, Inc., (RMD) for providing the effective active volume of the C7LYC crystal.

This project has received funding from the European Union's Horizon 2020 research and innovation programme under grant agreement No 654002, within the ENSAR2 project and JRA PASPAG.

## References

[1] C. M. Combes et al., Optical and scintillation properties of pure and $Ce^{3+}$-doped $Cs_2LiYCl_6$ and $Li_3YCl_6 : Ce^{3+}$ crystals, Journal of Luminescence B **82** (1999) pg. 299-305

[2] J. Glodo et al., Scintillation Properties of 1 Inch $Cs_2LiYCl_6$:Ce Crystals, IEEE Transactions on Nuclear Science **55**,3 (2008) pg. 1206

[3] J. Glodo et al., Development of $Cs_2LiYCl_6$ scintillator, Journal of Crystal Growth **379** (2013) pg. 73-78

[4] J. Glodo et al., Selected Properties of $Cs_2LiYCl_6$, $Cs_2LiLaCl_6$, and $Cs_2LiLaBr_6$ Scintillators, IEEE Transactions on Nuclear Science **58**,1 (2011) pg. 333-338

[5] A. Giaz et al., The C6LYC and C7LYC response to γ-rays, fast and thermal neutrons, Nuclear Instrumental and Methods in Physics Research A **810** (2016) pg. 132-139

[6] N. D'Olympia et al., Optimizing $Cs_2LiYCl_6$:Ce for fast neutron spectroscopy, Nuclear Instrumental and Methods in Physic Research A **694** (2012) pg. 140-146

[7] N. D'Olympia et al., Pulse-shape analysis of CLYC for thermal neutrons, fast neutrons, and gamma-rays, Nuclear Instrumental and Methods in Physic Research A **714** (2013) pg. 121-127

[8] M. B. Smith et al., Fast Neutron Spectroscopy Using $Cs_2LiYCl_6$:Ce (CLYC) Scintillator, IEEE Transactions on Nuclear Science **60**,2 (2013) pg. 885

[9] M. B. Smith et al., Fast neutron measurements using $Cs_2LiYCl_6$:Ce (CLYC) scintillator, Nuclear Instrumental and Methods in Physic Research A **784** (2014) pg. 162-167

[10] N. D'Olympia et al., Fast neutron response of $^6$Li-depleted CLYC detectors up to 20 MeV, Nuclear Instrumental and Methods in Physics Research A **763** (2014) pg. 433-441

[11] A. Giaz et al., Fast neutron measurements with $^7$Li and $^6$Li enriched CLYC scintillators, Nuclear Instrumental and Methods in Physic Research A **825** (2016) pg. 51

[12] R.S. Woolf et al., Response of the $^7$Li enriched $Cs_2LiYCl_6$:Ce (C7LYC) scintillators to 6-60 MeV neutrons, Nuclear Instrumental and Methods in Physic Research A **803** (2015) pg. 47-54

[13] B.S. McDonald et al., A wearable sensor based on CLYC scintillators, Nuclear Instrumental and Methods in Physics Research A **821** (2016) pg. 73-80

[14] J. Glodo et al., Integrated Neutron Detector for Handheld Systems, IEEE Transactions on Nuclear Science **60**, (2013)903

[15] C. Allwork et al., Neutron efficiency and Gamma rejection performances of CLYC and $^3$He alternative technologies IEEE Nuclear Science Symposium and Medical Imaging Conference (NSS/MIC) San Diego, CA (2015) pg. 1-7

[16] Lakshmi Soundara-Pandian et al., IEEE transactions on Nuclear Science **64**, (2017), 1744

[17] http://rmdinc.com/about-us/dynasil-corporate-profile/

[18] L. Lebreton et al., Rad. Prot. Dosimetry 2007, Vol. 126, pg. 3

[19] IAEA NDS/EXFOR <https://www-nds.iaea.org/exfor/endf.htm>